# The Research of the Real-time Detection and Recognition of Targets in Streetscape Videos

Liu Jian-min

**Abstract**

This study proposes a method for the real-time detection and recognition of targets in streetscape videos. The proposed method is based on separation confidence computation and scale synthesis optimization. We use the proposed method to detect and recognize targets in streetscape videos with high frame rates and high definition. Furthermore, we experimentally demonstrate that the accuracy and robustness of our proposed method are superior to those of conventional methods.

**1. Introduction**

Numerous Internet companies provide online streetscape image, mapping, and navigation services. All types of network media store large numbers of streetscape images and have gradually accumulated a massive volume of streetscape images and video data over time. The basic hardware conditions and data sources necessary for the automatic detection of related targets in videos are currently available. Moreover, accurate target detection is the basis of subsequent classification, target tracking, and behavioral analysis[1,2].

It is the key factor to improve the intelligence level of the monitoring system network that an automated streetscape videos target real-time detection recognition technique. By applying depth learning method to this field, multiple video sources and data sources are aggregated to realize intelligent detection and recognition for the related departments. It is very meaningful to provide information of accurate and real-time for judgement and decision.

We describe the status of research on machine-learning technology in the field of target detection and recognition below:The deep learning algorithm was first applied in image classification. Krizhevsky et al. proposed AlexNet, a deep convolution neural network that comprises eight learning layers. Specifically, AlexNet comprises five convolution layers and three fully connected layers. It achieved a top-5 classification error rate of as low as 15.3% when applied to classify images from the ImageNet dataset on the basis of dropout and random gradient descent[3]. In 2015, He et al. proposed the use of deep residual neural network to address the performance degradation caused by increasing the number of layers of the deep neural network to 152. To facilitate training, they used congruent mapping to directly connect the preceding output layer to the upper layer. The classification error rate of the deep residual neural network when used to classify images form the ImageNet dataset decreased to 3.6%, which is 5% less than the error rate achieved through human visual detection[4].

In 2013, Szegedy et al. proposed a target detection method based on convolution neural network and regression. This detection method yielded a mAP of 30.5% when applied on the VOC2007 dataset[5]. In 2014, Girshick et al. presented a region-based convolution neural network (RCNN), which transforms the target detection problem into a classification problem solvable by the convolution neural network. The mAP of the RCNN reached 58% when used to classify images from the VOC2007 dataset[6]. In 2015, Girshick et al. developed a region-based fast convolution neural network (FAST RCNN) that maps regions directly to the feature graph on the last convolution layer of the convolution neural network. FAST RCNN increased computational speed by transforming the target detection problem into a classification problem that is solvable by the convolution neural network. It yielded a mAP of 68% when used applied on the VOC2007 dataset[7].

In 2015, Girshick and He proposed a region-based faster convolution neural network (Faster-RCNN), which aggregated feature extraction, proposal extraction, bounding box regression, and classification to detect targets with quasi real-time speed. The mAP of VGG-based Faster-RCNN reached 76.4% when used to classify images from the VOC2007 dataset[8]. In 2016, Redmon developed a target detection algorithm that was based on the regression method. The algorithm achieved a mAP of 57.9% in the VOC 2012 dataset on the basis of end-to-end network from image input to output target location and detected category[9].

---

\* Liu jian-min.

Image classification and target detection with the deep learning neural network framework has become an important development model[10]. However, some methods for image detection and classification have been unable to keep up with the rapid development of intelligent monitoring, intelligent traffic, and other fields and the expansion of data sources for streetscape videos with high frame rate and definition and complex real scenes[11,12]. Therefore a new method needs to be studied for the real-time detection and recognition of targets in streetscape videos.

## 2. Method

As shown below，the proposed method is based on separation confidence computation and scale synthesis optimization. First, on the basis of generalization in transfer learning, we combine a fine-tuning method suitable for non-convex optimization and adaptive moment estimation in high-dimensional space. Then, we dynamically adjust the learning rates of parameters on the basis of first and second gradient moment estimations. We establish the framework and implementation steps of the proposed method by organically combining regular term super parameter generalization and hard-example mining technology. Through the organic combination of the above methods, We use the proposed method to detect and recognize targets in streetscape videos with high frame rates and high definition.

*2.1 Separation confidence computation*
In contrast to the selective target detection algorithm[13], image scanning generates rectangular selective boxes[14,15]. This behavior improves real-time performance. In image scanning, the video frame is divided into a fixed number of rectangular selection frames that may contain the object of interest. The corresponding target and classification probability are the calculated on the basis of the area encompassed by the rectangular selection frame. Thus, the calculation cost of this method is considerably reduced. The specific steps are as follows: First, the length and width of the video input image are divided into N parts, and the image is divided into N × N segments. The target object is predicted from the segment containing the center of the object. Videos of streetscapes, such as public squares, pedestrian streets, and dense traffic arteries, contain high densities of images of pedestrians and vehicles. To avoid missing valuable targets, N is set as 9. N is set as 9 in accordance with theoretical analysis, as shown in Fig.1. The algorithms are trained and tested independently, and the mAP[16] and detection rate are calculated.

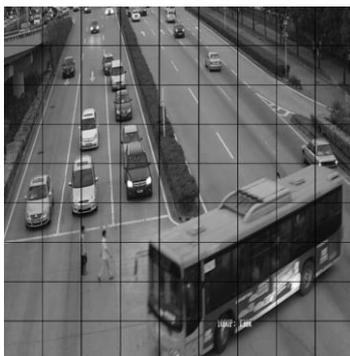

Fig.1 Horizontal–vertical division of a streetscape video image frame (N = 9)

To enhance generalization ability and to account for the different length–width ratios of different types of targets and the variations in the length–width ratios of the same type of targets, we designed six kinds of anchor selection boxes with different proportions. We specifically designed these boxes to account for pedestrians, nonmotor vehicles, motorcycles, cars, truck vehicles, and buses in images and to increase the probability that the anchor selection boxes contain the targets of interest. The length–width ratios of the six types of anchor selection boxes are designed as 1:1，1:2，2:1，2:2，2:4 and 4:2 (Fig.2).

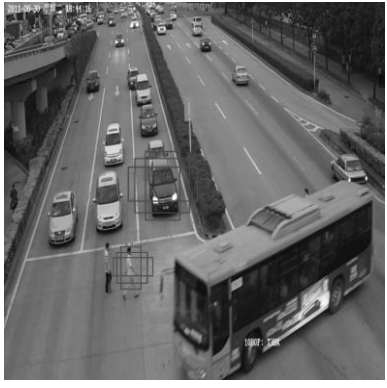

Fig.2 Examples of six selection boxes in image frames taken from a streetscape video

As shown in Fig.4, after each image frame of the streetscape video is processed, we set N = 9 and B = 6. Pedestrians, various nonmotor vehicles, and motor vehicles in the video are set as the detection targets. We set six categories, including pedestrians, nonmotor vehicles, motorcycles, cars, truck vehicles, and buses. The output is the detector.

Taking N = 9, the image is divided into 9 × 9 = 81 segments. Each segment is used to predict six categories with six anchor selection boxes. The six confidence levels are represented by 0, 1, 2, 3, 4, and 5. The coordinates of each anchor selection box are [a b w h], and each selection box has a confidence level. Each segment is used to predict 6 + 6 × 4 + 6 = 36 parameters. Each image frame from the video predicts 9 × 9 × 36 = 2916 parameters. Each segment contains five categories and six selection boxes. The six confidence levels are represented by 0, 1, 2, 3, 4, and 5. The coordinates of each selection box are [a b w h]. Each selection box has a confidence level. That is, each segment is used to predict 6 + 6 × 4 + 6 = 36 parameters. A total of 13 × 13 × 36 = 6084 parameters are predicted per image. In the above three cases, the coordinates of each selection box are [a b w h]. The coordinate [a b] is normalized to [0–1] by the corresponding separation relative to the deviation of each frame image, and [w h] is normalized to [0–1] on the basis of the width and height of the image.

*2.2 Competition of scale-based synthesis optimization*

In the detection of streetscape targets, such as traffic tools and pedestrians, two vectors are obtained: the predicted rectangular selection box that may contain the target, and the value obtained after the classifier is applied to the rectangular selection box that may contain the target. However, mutual inclusions or overlapping may exist, thus occluding objects in the rectangular selection box. Thus, the appropriate strategies for synthesis optimization must be designed to resolve the problem of occlusion. Nonmaximum suppression (NMS) screens elements and only allows the retention of the largest element[17] in an interval. This interval, in turn, contains two variables: the dimension and the range of the interval. Therefore, in this study, the scale-based synthesis optimization strategy is applied to identify the optimal rectangular selection box.

Given the influence of occlusion and position, low-scoring rectangular selection boxes may match with the relevant target. Therefore, although the initial rectangular selection box that surrounds each target is redundant, it can also be used to improve the accuracy of NMS before it is deleted. In this section, we discuss the method for the generation of the optimal rectangular selection box through NMS with scale-based optimization. When the rectangular selection box window $W_u$ with the highest score (its detection score is $S_u$) is fully included in the rectangular selection box $W_i$ with a low target score and large scale (its detection score $S_i$), and the detection scores $\frac{S_u - S_i}{S_u} < \lambda$ are met, $W_i$ is retained and $S_u = S_i$ is implemented. Otherwise, $W_i$ is directly deleted, and NMS is generated on the basis of the scale optimization strategy, which has strong robustness. Setting $\lambda \in [0.15, 0.2]$ considerably decreases the missed detection rate and false detection rate of the target rectangular selection box.

*2.3 Model generalization based on transfer learning*

Although streetscape videos are easily obtained, the number of mature datasets and the amount of data that can be used for streetscape object label training remain limited. The inappropriate selection of the initialization mode of the network model parameters is an important reason for model overfitting. The relative scarcity of accurately tagged streetscape image data is a major technical problem encountered in the development of detection models. To solve this problem, a model generalization technology based on transfer learning[18] is introduced to transfer the pretraining network model to the application of streetscape classification.

On the basis of transfer learning technology, under Class A Domain $D_A$, task $T_A$, and $T_A \in D_A$, the pretraining knowledge optimization model $M_B$ is transferred to the new application scene. We continue to use the network parameters and topology of the pretraining knowledge optimization model $M_B$, that is, we pretrain the algorithm with massive, universal image dataset. Then, distinct areas in small-scale streetscape datasets are fine-tuned to improve the classification accuracy of the proposed method for specific applications.

*2.4 Fine-tuning of stochastic optimization based on adaptive moment estimation*

Stochastic optimization based on adaptive moment estimation is a weight-adjustment method for the optimization of non-convex and high-dimensional space. The first moment estimation of the gradient is set to be $m_t$, and the second moment estimation of the gradient is set to be $n_t$. After continuously optimizing the learning rate by $m_t$ and $n_t$, they can be regarded as the estimation of expectation $E|g_t|$ $E|g_t^2|$. $\hat{m}_t$ $\hat{n}_t$ is the correction of $m_t$ $n_t$, which can be approximated as the unbiased estimation of the expectation. Its formula is given below:

$$\Delta \theta_t = -\frac{\hat{m}_t}{\epsilon + \sqrt{\hat{n}_t}} * \eta \qquad (1)$$

To achieve intensive resource demand, stochastic optimization based on adaptive moment estimation exploits the ability of the adaptive adjustment learning rate gradient method for high-performance computer sparse gradient and adaptive learning rate gradient method for high-performance response to unsteady objects. The corresponding parameters are autonomously optimized to match their learning rates. In this way, the floating range of the stochastic optimization learning rate of the adaptive moment estimation is limited to $-\frac{\hat{m}_t}{\varepsilon + \sqrt{\hat{n}_t}}$. The parameters under this constraint can be applied to non-convex optimization because of their stability, which is suitable for large datasets and high-dimensional space.

*2.5 Hard-example mining*

A large number of samples should be input during network training to ensure effective training. In addition, each sample has different effects on the fine-tuning result of the network. Some good samples aid training, whereas hard examples complicate training. However, if the training set contains an excessive proportion of good examples, the generalization ability and accuracy of the network will be affected. The addition of h numbers of hard examples in the training set can improve the effectiveness of the whole network but also increases computational time. After the completion of network training, new samples need to be added to the training set. We use Online Hard-example Mining[19] to improve efficiency and accuracy rate.

*2.6 Design of compound loss function*

We use a $9\times 9$ segment as an example to design a compound loss function. In reference to the regularization term λ, the probability and the error between the predictive rectangular selection box and the real value are weighted to generalize detection and recognition ability.

Input: $S_i(x_g, y_g, \omega_g, h_g)$ refers to the labeled target selection box, and $x_g, y_g$ refers to the central point.

Input: $S(x, y, \omega, h)$ refers to the preselection box of the target detection algorithm, and $x, y$ refers to the central point.

I: The number of segments in each image frame. Each segment contains six selection boxes (i = 0....max = N $\times$ N, 9 $\times$ 9 $\times$ 6).

$\lambda_{noobject}$ The weight of nontarget separation IoU loss for the gradient computation is set to 0.1 for optimization.

$\lambda$ is set to 0.2 to optimize the weight of the target separation IoU loss and classification loss for the gradient computation.

$\prod_i^{object}$ represents a selection box of the separation divided by each frame image, tab is i (i=0....485, 9×9×6) containing the concerned target.

$\prod_i^{noobject}$ represents a segment in the selection box divided by each frame image. The tab is i (i = 0....486, 9 ×9 ×6) and excludes the target output of concern.

Given that 1 out of 81 segments is required to calculate the probabilities of six categories, the original data of each input generate 486 probabilities through the network. The probability of a space point is set to $Pr(Object)$. Then, the target can be quantitatively measured. After 1 out of 81 segments is given by six categories of probabilities, the probability of whether the segment contains any target $Pr(Object)$ has to be quantified. Then, after each segment is divided into six categories, a single independent category of unconditional probability compounds the categorization probability of a target after identification, as shown in Eq. (2):

$$Pr(Car) = Pr(Object) * Pr(Car|Object) \qquad (2)$$

To summarize, the weighted linear combination of the independent loss function is used to optimize $Loss_{all}$, the total loss value generated, as shown in Eq.(3):

$$\begin{aligned}
Loss_{all} &= l_2\ loss + IoU\ loss + Class\ loss \\
&= \sum_{i=0}^{80} [\prod_i^{object} ((x_i - x_g)^2 + (y_i - y_g))^2] \\
&+ \sum_{i=0}^{80} (\prod_i^{object} [(\sqrt{\omega_i} - \sqrt{\omega_g})^2 + (\sqrt{h_i} - \sqrt{h_g}))^2] \\
&+ \lambda_{noobject} \sum_{i=0}^{80} \prod_i^{noobject} [\ln(\frac{S_i \cap S_g}{S_i \cup S_g - S_i \cap S_g})]^2 \\
&+ \lambda \sum_{i=0}^{80} \prod_i^{object} [\ln(\frac{S \cap S_g}{S_i \cup S_g - S_i \cap S_g})]^2 \\
&+ \lambda \sum_{i=0}^{80} \prod_i^{object} \sum_{c \in classifications} (p_i(c) - p_i(c_g))^2
\end{aligned} \qquad (3)$$

*2.7 Model design*

The designed model for the detection of targets in streetscape videos should achieve end-to-end image feature extraction and target detection and recognition. The training of the model are shown in Figs.5. We design the whole convolution network architecture to enhance the adaptability of input data and fully exploit the spatial information of the target.

1. A fine-tuning method for model generalization based on transfer learning and stochastic optimization for adaptive moment estimation is adopted.

With the initial min-batch gradient descent of the model, every time of the iterative learning rate lr is changed from 0.01 to 0.001. This learning rate attenuation strategy aims to learn new knowledge without completely forgetting old knowledge.

2. The anchor selection box is competitively optimized through scale-based synthesis optimization.

When the model is trained for a specific input image, if the boundary of the anchor selection box exceeds the image boundary, training loss should be unaffected by such an anchor selection box. The Loss value of this type of anchor selection box will be directly shielded. After shielding the anchor selection box that intersects with the image boundary, the remaining anchor selection boxes will overlap with each other at numerous regions. Therefore, the scale-based synthesis optimization strategy is used to select the anchor selection box for the development of competitive optimization with IoU > 0.7, and the anchor selection box that wins the competition is retained. Similarly, in the final output layer, at the confidence value >P and IoU > 0.7, each rectangular selection box is competitively optimized through the scale-based comprehensive optimization strategy, and the final winner is output.

3. The design of the whole convolution network architecture can enhance adaptability to the input data and fully exploit the spatial information of the target of concern. Enhancing data processing by using the target detection training dataset can appropriately expand the dataset size and reduce the over-fitting.

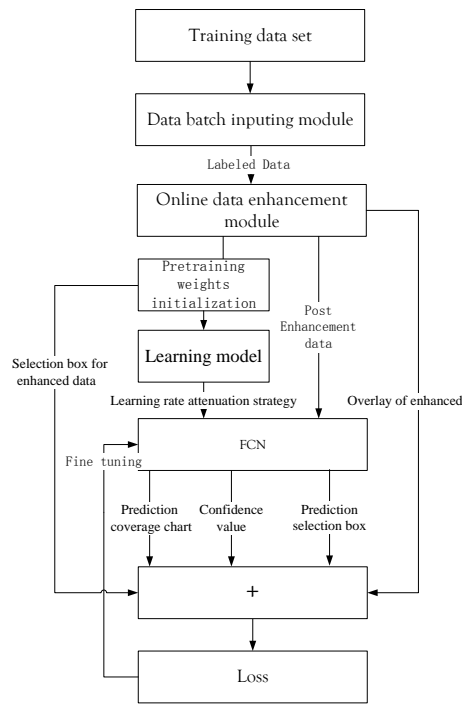

Fig.3 Training flow chart of the proposed method

## 3. Experiments

*3.1. Experimental setup*

The Pascal VOC 2007+2012 dataset[20], KITTI dataset[21], and Udacity dataset[22] discussed in Section 2.5 are adopted for pretraining and evaluating the proposed method for real-time target detection and recognition in streetscape videos. Sixteen streetscape videos with different resolutions are obtained from a traffic intersection monitoring terminal. Among these videos, nine are in D5 format. TargetLabelingInVideos tool is used to tag 319 video frames. The dataset is converted to the targeted fine-tuning training set with the same standard for transfer learning by a data preprocessing program, and 20% of data are reserved for testing data. We quantify the performance of the proposed method on the basis of the target detection comprehensive index mAP value[16]. This comprehensive evaluation index is obtained and calculated on the basis of recall, precision, and average precision (AP) and enables comprehensive measurement. To facilitate comparison with other methods, we define IOU > 0.5 as successful matching. The experimental platform is four-way 1080Ti and i7 7700K with the memory capacity of 4 TB.

*3.2 Analysis and discussion of experimental results*

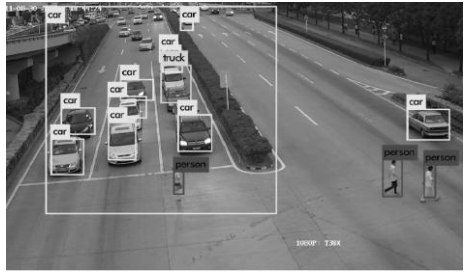

Fig.4 The proposed method for Streetscape video with D5 resolution 1 and instantaneous test output with N = 9

We set N = 9 segments and perform independent training and testing. The instantaneous test output for streetscape videos with N = 9 and D5 resolution is shown as Fig.4. The obtained mAP are shown in Figs.5. As shown in Tabs 1 and 2, the experiment involves 30 training cycles and a total of 95.6 h of training time.

Table 1 shows the comparison of the mAP values of this method with those of other detection methods. Tab 2 shows the comparison of the mAP and FPS of this method with those of other detection methods. mAP values tend to stabilize after the 25th training cycle when N = 9, 11, 13. During 25 training cycles, loss_bbox(train), loss_coverage(train), loss_bbox(val), and loss_coverage(val) values gradually converge and stabilize. N values do not significantly affect the training and convergence results. mAP is 90.9 and fps is 70 when N = 9. mAP is 91.74 and fps is 57 when N = 11. mAP is 91.56 and fps is 48. The above fps values are obtained from the detection of videos with D5 resolution.

When the N value is changed from 9 to 11–13, the mAP value of pedestrian detection increases from 91.1 to 93.3–93.5, and the mAP value of nonmotor vehicle detection changes from 90.5 to 92.4–92.55. Given the small size of pedestrians, nonmotor vehicles, and other targets, segmentation intensity is positively related with accuracy.

The classification accuracy of the proposed method for motorcycles, cars, and other medium-sized targets is 89.8 and 91.7 when N=9. The classification accuracy of the proposed method for motorcycles, cars, and other medium-sized targets is 90.2 and 91.6 when N=11. The classification accuracy of the proposed method for motorcycles, cars, and other medium-sized targets is 90 and 91.1 when N=13.

When N = 11, segmentation intensity is moderate, and the highest classification accuracy for motorcycles, cars, and other medium-sized targets is obtained. The classification accuracy of the proposed method for buses, trucks and other large targets is 91.5 and 91.3 when N = 9. The classification accuracy of the proposed method for buses, trucks and other large targets is 91.3 and 91.1 when N = 11. The classification accuracy of the proposed method for buses, trucks and other large targets is 90.7 and 90.7 when N =13. However, when N=9, the intensity of separation weakens, and the classification accuracy for buses, trucks and other large-sized targets increases.

As N is increased from 9 to 11–13, computational cost increases, whereas processing capacity decreases. Fps gradually decreases from 70 from 57–48. The processing capacity when N = 9 fully meets the requirement for target detection in streetscape videos shot by cameras with the high frame rate of 60 fps and can be also applied to multichannel videos and some platforms with insufficient processing capacity.

The processing capacity when N = 11 can fully meet the requirements for processing streetscape videos taken by a conventional 30 fps streetscape camera but is insufficient for processing streetscape videos shot by a 60 fps cameras. When N = 13, computational cost increases, and the processing capacity cannot meet the requirements for target detection from streetscape videos with high frame rates of 60 fps. Nevertheless, the detection accuracy for small and distant targets in streetscape videos increases. Thus, N can be set to 13 for the accurate real-time detection of small targets from streetscape videos with low frame rates.

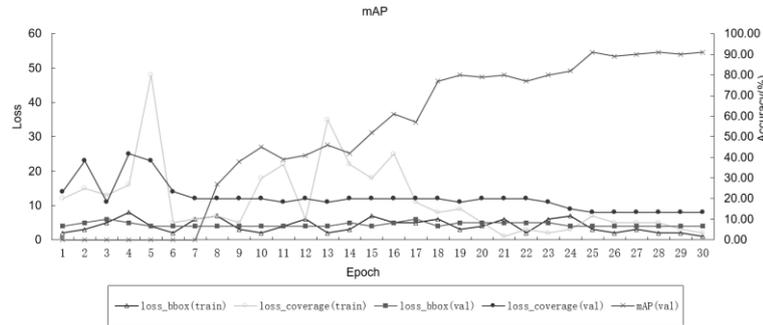

Fig.5 Training loss value and accuracy curves of the proposed method with 30 training cycles and N =9

Table 1 Comparison of the mAP values of this method

| Method | mAP | Person | Motor | Car | Bicycle | Bus | Truck |
|---|---|---|---|---|---|---|---|
| N=9, D5 resolution | 90.90 | 91.1 | 89.8 | 91.7 | 90.5 | 91.5 | 91.3 |

Table 2 Comparison of the mAP and FPS values of the proposed method with those of other detection methods

| Method | mAP | Training time(hours) | FPS(Testing) |
|---|---|---|---|
| N = 9, D5 resolution | 90.9 | 65.3 | 70 |

The method developed in this work integrates various methods for target detection. The proposed method yields a mAP of 90.9 and FPS of 70 when N = 9 for an input data source with D5 resolution. When N = 11, the proposed method obtains mAP and FPS values of 91.74 and 57, respectively. When N = 13, the proposed method obtains mAP and FPS values of 91.56 and 48, respectively. The detection accuracy and speed of the proposed method are superior to those of several classical methods, such as the Faster R-CNN ResNet, Faster R-CNN VGG-16, and YOLO v2 544×544 methods.

4. **Conclusion**

We proposed a real-time detection and recognition method based on separation confidence computation and scale synthesis optimization to address the problems encountered in target detection from streetscape videos with high frame rate and high definition. Moreover, we establish the framework and implementation steps of the proposed method. mAP and FPS measurements indicate that the performance of the proposed method has been improved by the combination of regular term superparameter generalization and hard-example mining technology. The proposed method meets the technical requirements for the detection of targets in streetscape videos with high frame rate and high definition.

**References**

1. Wu S, Chen D, Wang X. Moving target detection based on improved three frame difference and visual background extractor[C]// International Congress on Image and Signal Processing, Biomedical Engineering and Informatics. IEEE, 2018.
2. Shen X, Song Z, Fan H, et al. Data level moving target detection algorithm based on Bernoulli random finite set[J]. Iet Signal Processing, 2018.
3. Krizhevsky A, Sutskever I, Hinton G E. Imagenet classification with deep convolutional neuralnetworks[C].Advances in neural information processing systems. 2012:1097-1105.
4. He K, Zhang X, Ren S, etal. Deep residual learning for image recognition[J]. arXiv preprintarXiv:1512.03385, 2015.
5. Christian Szegedy, Alexander Toshev and Dumitru Erhan Deep Neural Networks for Object Detection[C]. Conference: Advances in Neural Information Processing Systems 26.Year: 2013,Pages: 2553--2561
6. R-CNN: Girshick R, Donahue J, Darrell T, et al. Rich feature hierarchies for accurate object detection and semantic

segmentation[C], CVPR, 2014.
7. Fast-RCNN: Girshick R. Fast R-CNN[C]. ICCV, 2015
8. Fater-RCNN: Ren S, He K, Girshick R, et al. Faster r-cnn: Towards real-time object detection with region proposal networks[C]. NIPS, 2016.
9. YOLO: Redmon J, Divvala S, Girshick R, et al. You Only Look Once: Unified, Real-Time Object Detection[J]. arXiv preprint arXiv:1506.02640, 2016.
10. Szegedy C, Liu W, Jia Y, et al. Going deeper with convolutions[C]. Proceedings of the IEEE Conference on Computer Vision and Pattern Recognition. 2015:1-9.
11. Li J, Wong H C, Lo S L, et al. Multiple Object Detection by Deformable Part-Based Model and R-CNN[J]. IEEE Signal Processing Letters, 2018, PP(99):1-1.
12. Dong P, Wang W. Better region proposals for pedestrian detection with R-CNN[C]// Visual Communications and Image Processing. IEEE, 2017:1-4.
13. Uijlings J R R, Sande KEAVD, Gevers T, et al. Selective Search for Object Recognition[J]. International Journal of Computer Vision, 2013, 104(2):154-171.
14. Redmon J, Farhadi A. YOLO9000: better, faster, stronger[J]. arXiv preprint arXiv:1612.08242, 2016.
15. Liu W, Anguelov D, Erhan D, et al. SSD: Single Shot MultiBox Detector[C].
European Conference on Computer Vision. Springer, Cham, 2016:21-37.
16. Zhu M. Recall, precision and average precision[J]. Department of Statistics and Actuarial Science, University of Waterloo, Waterloo, 2004, 2: 30.
17. Neubeck A, Van Gool L. Efficient non-maximum suppression[C]. Pattern Recognition, 2006. ICPR 2006. 18th International Conference on. IEEE, 2006, 3: 850-855.
18. Pan S J, Yang Q. A Survey on Transfer Learning[J]. IEEE Transactions on Knowledge & Data Engineering, 2010, 22(10):1345-1359.
19. Shrivastava A, Gupta A, Girshick R. Training region-based object detectors with online hard example mining[C]. Proceedings of the IEEE Conference on Computer Vision and Pattern Recognition. 2016: 761-769.
20. Everingham M, Van Gool L, Williams C K I, et al. The pascal visual object classes challenge 2012 (voc2012) results (2012)[C].URL http://www. pascal-network. org/challenges/VOC/.
21. Geiger A, Lenz P, Urtasun R. Are we ready for autonomous driving the kitti vision benchmark suite[C].Computer Vision and Pattern Recognition (CVPR), 2012 IEEE Conference on. IEEE, 2012: 3354-3361.
22. Udacity. Public driving dataset. https://www.udacity.com/self-driving-car, 2017. [EB/OL].(2017-3-07)[2017-9-21].
23. Russakovsky O, Deng J, Su H, et al. Imagenet large scale visual recognition challenge[J]. International Journal of Computer Vision, 2015, 115(3): 211-252.